# Enhancing speaker identification performance under the shouted talking condition using second-order circular hidden Markov models

Ismail Shahin

Electrical/Electronics and Computer Engineering Department

University of Sharjah

P. O. Box  27272

Sharjah, United Arab Emirates

Tel: (971) 6 5050967, Fax: (971) 6 5050877

E-mail: ismail@sharjah.ac.ae

**Abstract**

It is known that the performance of speaker identification systems is high under the neutral talking condition; however, the performance deteriorates under the shouted talking condition. In this paper, second-order circular hidden Markov models (CHMM2s) have been proposed and implemented to enhance the performance of isolated-word text-dependent speaker identification systems under the shouted talking condition. Our results show that CHMM2s significantly improve speaker identification performance under such a condition compared to the first-order left-to-right hidden Markov models (LTRHMM1s), second-order left-to-right hidden Markov models (LTRHMM2s), and the first-order circular hidden Markov models (CHMM1s). Under the shouted talking condition, our results show that the average speaker identification performance is 23% based on LTRHMM1s, 59% based on LTRHMM2s, and 60% based on CHMM1s. On the other hand, the average speaker identification performance under the same talking condition based on CHMM2s is 72%.

## 1. Introduction

The goal of speaker identification systems is to decide which voice model from a known set of voice models best characterizes a speaker. The applications of speaker identification may be required in criminal investigations to determine the suspected persons produced the voice recorded at the scene of the crime, it may also be required in civil cases or for the media. These cases include calls to radio stations, local or other government authorities, insurance companies, or recorded conversations, and many other applications.

Speaker identification systems typically operate in one of two cases: text-dependent or text-independent case. In the text-dependent case, utterances of the same text are used for both training and testing (recognition). In the text-independent case, training and testing involve utterances from different texts. The process of speaker identification can be divided into two categories: "open set" and "closed set". In the "open set" category, a reference model for an unknown speaker may not exist; whereas, in the "closed set" category, a reference model for an unknown speaker should be available.

## 2. Motivation

In the last three decades, hidden Markov models (HMMs) have become one of the most successful and broadly used modeling techniques in the fields of speech recognition and speaker recognition (Juang and Rabiner, 1991). HMMs are powerful techniques in optimizing the parameters that are used in modeling speech signals. This optimization reduces the computational complexity in the decoding procedure and improves the recognition accuracy (Huang et al., 1990).

Most of the publications in the areas of speech recognition and speaker recognition focus on speech under the neutral talking condition and few publications focus on speech under stressful talking conditions. The neutral talking condition can be defined as the talking condition in which speech is produced assuming that the speaker is in a "quiet room" with no task obligations. Stressful talking conditions can be defined as talking conditions that cause a speaker to vary his/her production of speech from the neutral talking condition.

Some talking conditions are designed to simulate speech produced by different speakers under real stressful talking conditions. Hansen, Cummings, and Clements used SUSAS (Speech Under Simulated and Actual Stress) speech database in which eight talking conditions are used to simulate speech produced under real stressful talking conditions and three real talking conditions (Bou-Ghazale and Hansen, 2000; Cummings and Clements, 1995; Zhou et al., 2001). The eight talking conditions are: neutral, loud, soft, angry, fast, slow, clear, and question. The three talking conditions are: 50% task, 70% task, and Lombard. Chen used six talking conditions to simulate speech under real stressful talking conditions (Chen, 1988). These conditions are: neutral, fast, loud, Lombard, soft, and shouted.

Very few publications that focus on speech under the stressful talking conditions consider studying speech under the shouted talking condition (Cairns and Hansen, 1994; Chen, 1988; Hansen, 1996). Therefore, the number of references in the areas of speech recognition and speaker recognition under the shouted talking condition is limited. The shouted talking condition can be defined as when a speaker shouts, his/her object is to produce a very loud acoustic signal, either to increase its range (distance) of transmission or its ratio to background noise.

Speaker identification systems under the shouted talking condition can be used in the applications of talking condition identification systems. Talking condition identification systems can be used in telecommunications, military applications, medical applications, and law enforcement. In telecommunications, talking condition identification systems can be used to: improve the telephone-based speech recognition performance, route 911 (in U.S.A.) or 999 (in U.A.E.) emergency call services for high priority emergency calls, and assess a caller's emotional state for telephone response services. The integration of speech recognition technology has already been seen in many military voice communication and control applications. Such applications involve stressful environments such as aircraft cockpits and military peacekeeping (Hansen et al., 2000). Talking condition identification systems can be used also in medical applications where computerized stress classification and assessment techniques can be employed by psychiatrists to aid in quantitative objective assessment of patients who undergo evaluation. Finally, talking condition identification systems can be employed in forensic speech analysis by law enforcement to assess the state of telephone callers or as an aid in suspect interviews.

Researchers are incorporating emotional capabilities into speech synthesis programs, hoping to enable computers that can communicate emotionally with users through expressive vocal signals such as laughter, sighing, or sad tones of voice. IBM is set to release a new Expressive Text-to-Speech Engine for commercial use that will deliver spoken information in the appropriate tone, and also include lifelike capabilities such as the ability to clear its throat, cough, and pause for breath (Stroh, 2004). AT&T Lab is developing the opposite technology, software that can detect users' emotional state; voice-response systems equipped with this software would be able to prioritize calls according to the person's state of agitation, for example (Stroh, 2004).

In the last three decades, the majority of the work performed in the fields of speech recognition and speaker recognition on HMMs has been done using LTRHMM1s (Chen, 1988; Dai, 1995; Juang and Rabiner, 1991; Rabiner, 1989). Despite the success of using LTRHMM1s under the neutral talking condition, they yield low recognition performance under the shouted talking condition.

In this paper, we show that using CHMM2s in the training and testing phases of isolated-word text-dependent speaker identification systems outperforms each of LTRHMM1s, LTRHMM2s, and CHMM1s under the shouted talking condition.

Our work in this paper differs from the work in (Mari et al., 1996; Mari et al., 1997; Zheng and Yuan, 1988) is that our work focuses on enhancing the performance of isolated-word text-dependent speaker identification systems under the shouted talking condition based on CHMM2s. The work in (Mari et al., 1996; Mari et al., 1997) focuses on describing a connected word recognition system under the neutral talking condition based on HMM2s. The work in (Mari et al., 1996; Mari et al., 1997) shows that the recognition performance based on HMM2s yields better results than that based on HMM1s. The work in (Zheng and Yuan, 1988) focuses on enhancing speaker identification performance under the neutral talking condition based on CHMMs.

## 3. Brief overview of left-to-right hidden Markov models

HMMs use Markov chain to model the changing statistical characteristics that exist in the actual observations of speech signals. HMMs are double stochastic process where there is an unobservable Markov chain defined by a state transition matrix, and where each state of the Markov chain is associated with either a discrete output probability distribution (discrete HMMs) or a continuous output probability density function (continuous HMMs) (Huang et al., 1990).

In the left-to-right (LTR) models, backward transitions are not allowed, and the models progress through the states in a left-to-right way. In these models, the state transition coefficients have the property of,

$$a_{ij} = 0 \quad \text{when} \quad j < i \tag{1}$$

This means that no backward transitions in time are allowed. In these models, the transition from a lower state to a higher state is the only transition that is allowed. Fig. 1 illustrates LTR models (Rabiner and Juang, 1983).

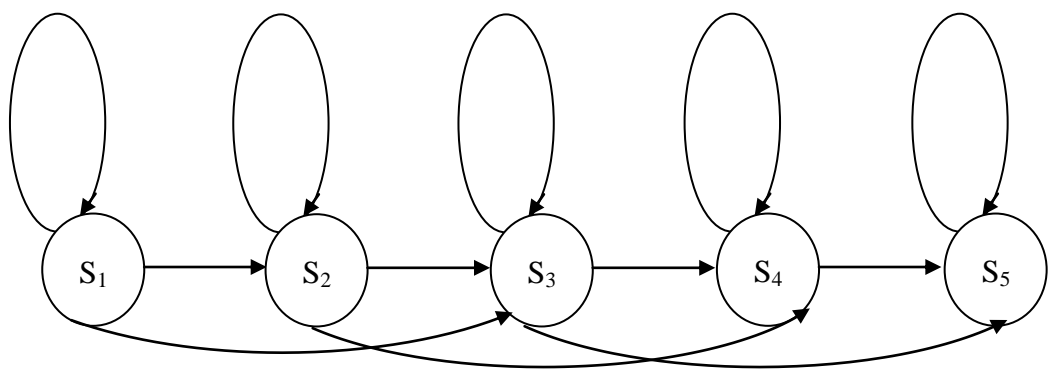

Fig. 1. 5-state left-to-right hidden Markov models with state skipping

The initial state probabilities are defined as,

$$\pi_i = \begin{cases} 0 & i \neq 1 \\ 1 & i = 1 \end{cases} \qquad (2)$$

The maximum reestimates of LTRHMMs parameters, $a_{ij}$ and $b_{jk}$, can be computed recursively using the Baum-Welch algorithm (Levinson et al., 1983),

$$a_{ij}^{n+1} = \frac{\sum_{t=1}^{T-1} \alpha_t(i) a_{ij}^n b_j^n(o_{t+1}) \beta_{t+1}(j)}{\sum_{t=1}^{T-1} \alpha_t(i) \beta_t(i)} \qquad i,j = 1,2,...,N \qquad (3)$$

where $\alpha_t(i)$ and $\beta_t(i)$ are respectively the forward and backward probabilities of producing the observation vector $O$ and can be calculated as,

$$\alpha_t(j) = \left\{ \sum_{i=1}^{N} \alpha_{t-1}(i) a_{ij} \right\} b_j(O_t) \qquad T-1 \geq t \geq 1 \qquad (4)$$

$$\beta_t(i) = \sum_{j=1}^{N} a_{ij} b_j(O_{t+1}) \beta_{t+1}(j) \qquad T-1 \geq t \geq 1 \qquad (5)$$

The parameter $b_{jk}$ can be reestimated using the segmental k-means algorithm which is used in the training procedure to optimally estimate and adjust the model parameters in building the reference models.

The probability of generating the observation vector $O$ can be computed using the Baum algorithm as,

$$P = \sum_{i=1}^{N} \sum_{j=1}^{N} \alpha_t(i) a_{ij} b_j(O_{t+1}) \beta_{t+1}(i) \qquad T \geq t \geq 1 \qquad (6)$$

More details about LTRHMMs can be found in many references (Dai, 1995; Juang and Rabiner, 1985; Juang and Rabiner, 1991; Levinson et al., 1983; Rabiner and Juang, 1983; Rabiner, 1989).

### 4.1 Second order hidden Markov models

HMM1s have been used in the training and testing phases of the vast majority of the work in the areas of speech recognition and speaker recognition (Chen, 1988; Dai, 1995; Juang and Rabiner, 1991; Rabiner, 1989). The recognition performance based on HMM1s is high under the neutral talking condition; however, the performance is degraded sharply under the shouted talking condition (Chen, 1988; Shahin and Botros, 1998a, 1998b).

New models called HMM2s were introduced and implemented under the neutral talking condition by Mari, Fohr, and Junqua (Mari et al., 1996). These models have shown to improve the performance of isolated-word text-dependent speaker identification systems under the shouted talking condition (Shahin, in press).

In HMM1s, the underlying state sequence is a first-order Markov chain where the stochastic process is specified by a 2-D matrix of a priori transition probabilities ($a_{ij}$) between states $s_i$ and $s_j$ where $a_{ij}$ are given as,

$$a_{ij} = \text{Prob}(q_t = s_j | q_{t-1} = s_i) \qquad (7)$$

In HMM2s, the underlying state sequence is a second-order Markov chain where the stochastic process is specified by a 3-D matrix ($a_{ijk}$). Therefore, the transition probabilities in HMM2s are given as (Mari et al., 1997),

$$a_{ijk} = \text{Prob}(q_t = s_k | q_{t-1} = s_j, q_{t-2} = s_i) \tag{8}$$

with the constraints,

$$\sum_{k=1}^{N} a_{ijk} = 1 \qquad N \geq i, j \geq 1$$

The probability of the state sequence, $Q \triangleq q_1, q_2, ..., q_T$, is defined as,

$$\text{Prob}(Q) = \Psi_{q_1} a_{q_1 q_2} \prod_{t=3}^{T} a_{q_{t-2} q_{t-1} q_t} \tag{9}$$

where $\Psi_i$ is the probability of a state $s_i$ at time $t = 1$, $a_{ij}$ is the probability of the transition from a state $s_i$ to a state $s_j$ at time $t = 2$.

Each state $s_i$ is associated with a mixture of Gaussian distributions,

$$b_i(O_t) \triangleq \sum_{m=1}^{M} c_{im} N(O_t; \mu_{im}, \Sigma_{im}), \qquad \text{with} \sum_{m=1}^{M} c_{im} = 1 \tag{10}$$

where the vector $O_t$ is the input vector at time $t$.

Given a sequence of observed vectors, $O \triangleq O_1, O_2, ..., O_T$, the joint state-output probability is defined as,

$$\text{Prob}(Q, O | \lambda) = \Psi_{q_1} b_{q_1}(O_1) a_{q_1 q_2} b_{q_2}(O_2) \prod_{t=3}^{T} a_{q_{t-2} q_{t-1} q_t} b_{q_t}(O_t) \tag{11}$$

## 4.2 Extended Viterbi and Baum-Welch algorithms

The most likely state sequence can be found by using the probability of the partial alignment ending at a transition $(s_j, s_k)$ at times $(t-1, t)$,

$$\delta_t(j,k) \triangleq \text{Prob}(q_1,...,q_{t-1}=s_j, q_t=s_k, O_1, O_2,..., O_t | \lambda) \quad T \geq t \geq 2, N \geq j, k \geq 1 \quad (12)$$

Recursive computation is given by,

$$\delta_t(j,k) = \max_{N \geq i \geq 1} \{\delta_{t-1}(i,j) \cdot a_{ijk}\} \cdot b_k(O_t) \quad T \geq t \geq 3, N \geq j, k \geq 1 \quad (13)$$

The forward function $\alpha_t(j,k)$ defines the probability of the partial observation sequence, $O_1, O_2, \ldots, O_t$, and the transition $(s_j, s_k)$ between times $t-1$ and $t$ is given by,

$$\alpha_t(j,k) \triangleq \text{Prob}(O_1,..., O_t, q_{t-1}=s_j, q_t=s_k | \lambda) \quad T \geq t \geq 2, N \geq j, k \geq 1 \quad (14)$$

$\alpha_t(j,k)$ can be computed from the two transitions: $(s_i, s_j)$ and $(s_j, s_k)$ between states $s_i$ and $s_k$ as,

$$\alpha_{t+1}(j,k) = \sum_{i=1}^{N} \alpha_t(i,j) \cdot a_{ijk} \cdot b_k(O_{t+1}) \quad T-1 \geq t \geq 2, N \geq j, k \geq 1 \quad (15)$$

The backward function $\beta_t(i, j)$ can be expressed as,

$$\beta_t(i,j) \triangleq \text{Prob}(O_{t+1},..., O_T | q_{t-1}=s_i, q_t=s_j, \lambda) \quad T-1 \geq t \geq 2, N \geq i, j \geq 1 \quad (16)$$

where $\beta_t(i,j)$ is defined as the probability of the partial observation sequence from $t+1$ to $T$, given the model $\lambda$ and the transition $(s_i, s_j)$ between times $t-1$ and $t$.

## 5. Circular hidden Markov models

Most of the work performed in the last three decades in the fields of speech recognition and speaker recognition using HMMs has been done using LTRHMMs (Chen, 1988; Dai, 1995; Juang and Rabiner, 1991; Rabiner, 1989). LTRHMMs yield high speaker identification performance under the neutral talking condition; however, the performance is deteriorated sharply under the shouted talking condition (Chen, 1988; Shahin and Botros, 1998a, 1998b).

New models called CHMMs were introduced and used by Zheng and Yuan under the neutral talking condition (Zheng and Yuan, 1988). CHMMs are considered as another special class of HMMs. These models yield high speaker identification performance under the neutral talking condition (Shahin, 2004). The structure of CHMMs is shown in Fig. 2. CHMMs have the following properties (Zheng and Yuan, 1988):

i. The underlying Markov chain has no final or absorbing state. Therefore, the corresponding HMMs can be trained by as long training sequence as desired.

ii. Once the Markov chain leaves a state, that state can be revisited only at the next time.

In CHMMs, the state transition coefficients have the property of,

$$a_{ij} = a_{ji} \qquad i, j = 1,2,\ldots,N \qquad (17)$$

Therefore, the state transition probability matrix **A** is a symmetrical matrix.

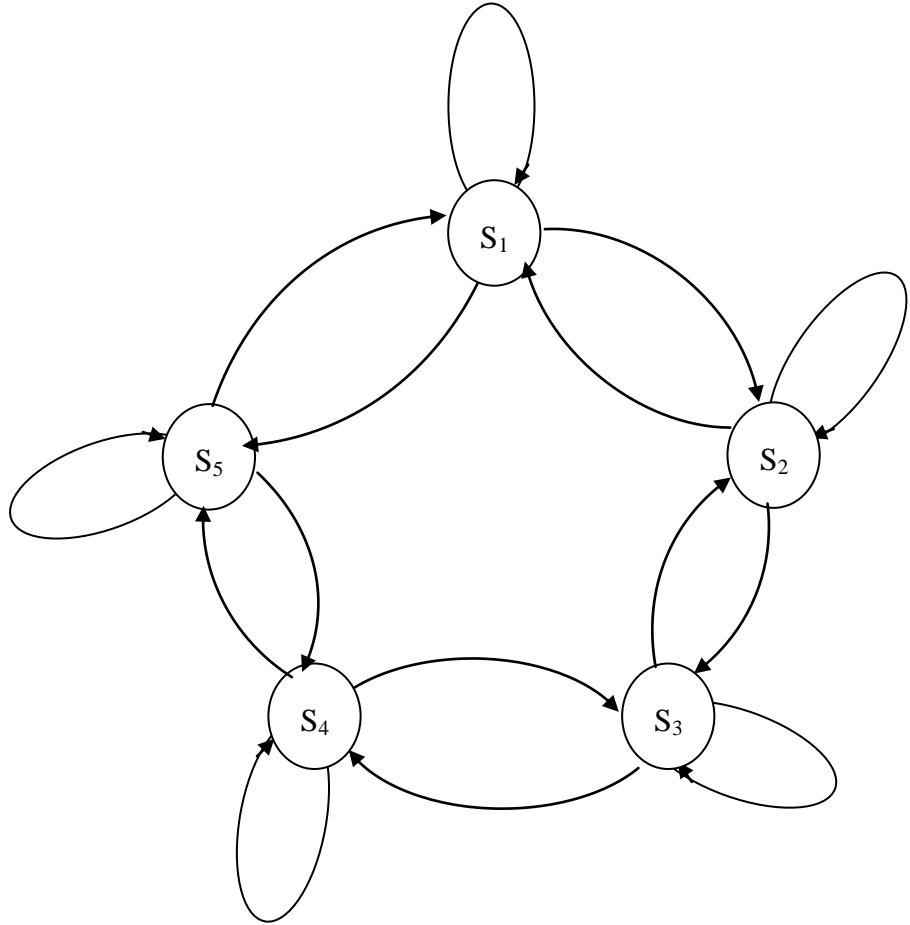

Fig. 2. 5-state circular hidden Markov models

The initial elements of the parameters in the training phase of CHMMs are chosen to be (Zheng and Yuan, 1988),

$$v(i) = \frac{1}{N} \qquad N \geq i \geq 1 \qquad (18)$$

where v(i) is the initial element of the probability of an initial state distribution.

$$\alpha_1(i) = v(i) b_i(O_1) \qquad N \geq i \geq 1 \qquad (19)$$

where $\alpha_1(i)$ is the initial element of the forward probability of producing the observation $O_1$.

$$a_{ij}^1 = \begin{cases} \dfrac{1}{3} & i=1, j=1,2,...,N \\ \\ \dfrac{1}{3} & N-1 \geq i \geq 2, i+1 \geq j \geq i-1 \\ \\ \dfrac{1}{3} & i=N, j=1,2,...,N \\ \\ 0 & \text{otherwise} \end{cases} \quad (20)$$

where $a_{ij}^1$ is the initial element of the transition probability from a state $S_i$ to a state $S_j$.

$$b_{jk}^1 = \frac{1}{M} \qquad N \geq j \geq 1, M \geq k \geq 1 \quad (21)$$

where $b_{jk}^1$ is the initial element of the observation symbol probability and M is the number of observation symbols.

$$\beta_T(j) = \frac{1}{N} \qquad N \geq j \geq 1 \quad (22)$$

where $\beta_T(j)$ is the initial element of the backward probability of producing the observation $O_T$.

## 6. Second-order circular hidden Markov models

New models called CHMM2s have been proposed and used in this work to enhance the performance of isolated-word text-dependent speaker identification systems under the shouted talking condition. This is the first known investigation into CHMM2s evaluated under the shouted talking condition for speaker identification systems. CHMM2s are considered as another special class of HMMs. CHMM2s are superiors over each of LTRHMM1s, LTRHMM2s, and

CHMM1s. This is because CHMM2s possess the characteristics of both CHMMs and HMM2s:

1. The underlying state sequence in HMM2s is a second-order Markov chain where the stochastic process is specified by a 3-D matrix because in these models the state-transition probability at time $t+1$ depends on the states of the Markov chain at times $t$ and $t-1$. On the other hand, the underlying state sequence in HMM1s is a first-order Markov chain where the stochastic process is specified by a 2-D matrix because in these models it is assumed that the state-transition probability at time $t+1$ depends only on the state of the Markov chain at time $t$. Hence, the stochastic process that is specified by a 3-D matrix gives more accurate speaker identification performance than that specified by a 2-D matrix (Shahin, 2005).

2. The Markov chain in CHMMs is more powerful and efficient in modeling the changing statistical characteristics that exist in the actual observations of the speech signals than that in LTRHMMs.

3. In LTRHMMs, the absorbing state governs the fact that the rest of a single observation sequence provides no further information about earlier states once the underlying Markov chain reaches the absorbing state. In speaker identification systems, it is true that a Markov chain should be able to revisit the earlier states because the states of HMMs reflect the vocal organic configuration of the speaker. Therefore, the vocal organic configuration of the speaker is reflected to states more appropriately and more conveniently using CHMMs than that using LTRHMMs. Therefore, it is inconvenient to utilize LTRHMMs having one absorbing state for speaker identification systems (Shahin, 2004).

The initial elements of the parameters in the training phase of CHMM2s are chosen to be,

$$v_k(i) = \frac{1}{N} \qquad N \geq i, k \geq 1 \qquad (23)$$

where $v_k(i)$ is the initial element of the probability of an initial state distribution.

$$\alpha_1(i,k) = v_k(i) b_{ki}(O_1) \qquad N \geq i, k \geq 1 \qquad (24)$$

where $\alpha_1(i,k)$ is the initial element of the forward probability of producing the observation $O_1$.

$$a^1_{ijk} = \begin{cases} \frac{1}{3} & i=1, j,k=1,2,...,N \\ \frac{1}{3} & N-1 \geq i \geq 2, i+1 \geq j \geq i-1, N \geq k \geq 1 \\ \frac{1}{3} & i=N, j,k=1,2,...,N \\ 0 & \text{otherwise} \end{cases} \qquad (25)$$

where $a^1_{ijk}$ is the initial element of $a_{ijk}$.

$$b^1_{ijk} = \frac{1}{M} \qquad N \geq j, k \geq 1, M \geq i \geq 1 \qquad (26)$$

where $b^1_{ijk}$ is the initial element of the observation symbol probability and M is the number of observation symbols.

$$\beta_T(j,k) = \frac{1}{N} \qquad N \geq j, k \geq 1 \qquad (27)$$

where $\beta_T(j,k)$ is the initial element of the backward probability of producing the observation $O_T$.

## 7. Speech database

Our speech database in this work is composed of 20 adult native American male speakers and 20 adult native American female speakers. Each speaker utters the same 10 different isolated words where each word is uttered 9 times (9 utterances per word) under each of the neutral and shouted talking conditions. These words are: alphabet, eat, fix, meat, nine, order, processing, school, six, yahoo. The length of these words ranges from 1 to 3 seconds.

Our speech database was captured by a speech acquisition board using a 10-bit linear coding A/D converter and sampled at a sampling rate of 8 kHz. Our database was a 10-bit per sample linear data. Each signal of the neutral and shouted talking conditions was applied to a high emphasis filter, $H(z)=1-0.95z^{-1}$. Each emphasized speech signal was applied every 10 ms to a 30-ms Hamming window. *12*th order linear prediction coefficients (LPCs) were extracted from each frame by the autocorrelation method. The *12*th order LPCs were then transformed into *12*th order linear prediction cepstral coefficients (LPCCs).

The LPCC feature analysis was used to form the observation vectors in each of LTRHMM1s, LTRHMM2s, CHMM1s, and CHMM2s. The number of states, $N$, was 5. The number of mixture components, $M$, was 5 per state, with a continuous mixture observation density was selected for each of LTRHMM1s, LTRHMM2s, CHMM1s, and CHMM2s. Our speech database was divided into training data under the neutral talking condition and test data under each of the neutral and shouted talking conditions. Our speech database in this work was a "closed set".

In the training session, one reference model is derived using 5 of the 9 utterances per the same speaker per the same word under the neutral talking condition. Training in this session has been done separately based on each of LTRHMM1s, LTRHMM2s, CHMM1s, and CHMM2s. Training of the models in each training session uses the forward-backward algorithm.

In the testing (identification) session, each one of the 40 speakers uses 4 of the 9 utterances per the same word (text-dependent) under the neutral talking condition and 9 utterances per the same word under the shouted talking condition. This session has been done separately based on each of LTRHMM1s, LTRHMM2s, CHMM1s, and CHMM2s. Identification in each testing session uses the Viterbi decoding algorithm. Our speech database is summarized in Table 1.

## 8. Results

Computing the probability of generating an utterance, the model with the highest probability is chosen as the output of the speaker identification system under each of the neutral and shouted talking conditions based on each of LTRHMM1s, LTRHMM2s, CHMM1s, and CHMM2s.

Table 1

Speech database under each of the neutral and shouted talking conditions

| Models | Session | Total number of utterances under the neutral talking condition | Total number of utterances under the shouted talking condition |
|---|---|---|---|
| LTRHMM1s | Training | 1 000 male utterances<br>1 000 female utterances | 0 male utterance<br>0 female utterance |
| | Testing | 800 male utterances<br>800 female utterances | 1800 male utterances<br>1800 female utterances |
| LTRHMM2s | Training | 1 000 male utterances<br>1 000 female utterances | 0 male utterance<br>0 female utterance |
| | Testing | 800 male utterances<br>800 female utterances | 1800 male utterances<br>1800 female utterances |
| CHMM1s | Training | 1 000 male utterances<br>1 000 female utterances | 0 male utterance<br>0 female utterance |
| | Testing | 800 male utterances<br>800 female utterances | 1800 male utterances<br>1800 female utterances |
| CHMM2s | Training | 1 000 male utterances<br>1 000 female utterances | 0 male utterance<br>0 female utterance |
| | Testing | 800 male utterances<br>800 female utterances | 1800 male utterances<br>1800 female utterances |

Table 2 summarizes the results of the speaker identification performance for 20 male speakers, 20 female speakers, and their averages under each of the neutral and shouted talking conditions based on each of LTRHMM1s, LTRHMM2s, CHMM1s, and CHMM2s.

Table 2

Speaker identification performance under each of the neutral and shouted talking conditions based on each of LTRHMM1s, LTRHMM2s, CHMM1s, and CHMM2s

| Models | Gender | Neutral | Shouted |
|---|---|---|---|
| LTRHMM1s | Males | 89% | 21% |
| | Females | 91% | 25% |
| | Average | 90% | 23% |
| LTRHMM2s | Males | 92% | 57% |
| | Females | 96% | 61% |
| | Average | 94% | 59% |
| CHMM1s | Males | 91% | 59% |
| | Females | 93% | 61% |
| | Average | 92% | 60% |
| CHMM2s | Males | 94% | 71% |
| | Females | 97% | 73% |
| | Average | 95.5% | 72% |

## 9. Discussion and conclusions

This work is based on an isolated-word text-dependent second-order circular hidden Markov models speaker identifier trained by speech uttered under the neutral talking condition and tested by speech uttered under each of the neutral and shouted talking conditions. This is the first known investigation into CHMM2s evaluated under the shouted talking condition for a speaker identification system.

Our work shows that training and testing an isolated-word text-dependent speaker identification system based on using CHMM2s significantly enhance the speaker identification performance under each of the neutral and shouted talking conditions compared to that based on using each of LTRHMM1s, LTRHMM2s, and CHMM1s. The average improvement rate of implementing CHMM2s over each of LTRHMM1s, LTRHMM2s, and CHMM1s under each of the neutral and shouted talking conditions is summarized in Table 3.

Table 3

Average improvement rate of implementing CHMM2s over each of LTRHMM1s, LTRHMM2s, and CHMM1s under each of the neutral and shouted talking conditions

| Models | Average improvement rate under the neutral talking condition | Average improvement rate under the shouted talking condition |
|---|---|---|
| LTRHMM1s | 6.1% | 213.0% |
| LTRHMM2s | 1.6% | 22.0% |
| CHMM1s | 2.7% | 17.1% |

It is evident from Table 3 that CHMM2s are superiors over each of LTRHMM1s, LTRHMM2s, and CHMM1s under each of the neutral and shouted talking conditions. Table 2 shows that under the shouted talking condition:

1. The average improvement rate of implementing LTRHMM2s over LTRHMM1s is 156.5%, which is a significant improvement of speaker identification performance.

2. The average improvement rate of implementing CHMM1s over LTRHMM1s is 160.9%, which is a significant improvement of speaker identification performance.

3. The average improvement rate of implementing CHMM2s over LTRHMM1s is 213%, which is a very significant improvement of speaker identification performance.

It is evident from the previous 3 points and Table 3 that implementing CHMM2s under the shouted talking condition yields a very significant improvement of speaker identification performance compared to each of LTRHMM1s, LTRHMM2s, and CHMM1s. This is because CHMM2s possess the characteristics of both CHMMs and HMM2s as were discussed in Section 6.

Table 3 shows that the speaker identification performance under the neutral talking condition has been improved significantly based on implementing CHMM2s compared to that based on implementing LTRHMM1s; the average improvement rate of implementing CHMM2s over LTRHMM1s under such a talking condition is 6.1%. Comparing the average improvement rate of implementing CHMM2s over LTRHMM1s under the shouted talking condition (213%) with that under the neutral talking condition (6.1%), it is evident that the average improvement rate of implementing CHMM2s over LTRHMM1s under the neutral talking condition is much less than that under the shouted talking condition. The reason is that it is known that LTRHMM1s are powerful and efficient models under the neutral talking condition (Chen, 1988; Dai, 1995;

Haung et al., 1990; Levinson et al., 1983); however, they are inefficient models under the shouted talking condition (Chen, 1988; Shahin and Botros, 1998a; Shahin, 2005).

An experiment has been conducted under each of the neutral and shouted talking conditions to compare the speaker identification performance based on implementing CHMM2s with that based on implementing each of LTRHMM1s, LTRHMM2s, and CHMM1s using the stress compensation technique.

It is well known that the spectral tilt exhibits a large variation when the speaker talks under the shouted talking condition (Chen, 1988). Such a variation usually contaminates the distance measure and it is considered as one of the most significant causes of degradation in the speaker identification performance. One of the stress compensation techniques that removes the spectral tilt and enhances the speaker identification performance is the cepstral mean subtraction technique (Shahin and Botros, 1998a).

Table 4 summarizes the results of the speaker identification performance under each of the neutral and shouted talking conditions for the 20 male speakers, 20 female speakers, and their averages based on implementing each of LTRHMM1s, LTRHMM2s, and CHMM1s using the cepstral mean subtraction technique. Comparing Table 2 with Table 4, it is apparent that CHMM2s are superiors over each of LTRHMM1s, LTRHMM2s, and CHMM1s using the cepstral mean subtraction technique under each of the neutral and shouted talking conditions.

A naïve implementation of the recursion for the computation of $\alpha$ and $\beta$ in each of CHMM2s and LTRHMM2s requires on the order of $N^3T$ operations, compared with $N^2T$ operations in each of CHMM1s and LTRHMM1s. Therefore, it is required more memory space in each of CHMM2s and LTRHMM2s than that in each of CHMM1s and LTRHMM1s.

Table 4

Speaker identification performance under each of the neutral and shouted talking conditions based on implementing each of LTRHMM1s, LTRHMM2s, and CHMM1s using the cepstral mean subtraction technique

| Models | Gender | Neutral | Shouted |
|---|---|---|---|
| LTRHMM1s | Males | 89% | 39% |
| | Females | 91% | 41% |
| | Average | 90% | 40% |
| LTRHMM2s | Males | 93% | 63% |
| | Females | 97% | 65% |
| | Average | 95% | 64% |
| CHMM1s | Males | 92% | 62% |
| | Females | 94% | 64% |
| | Average | 93% | 63% |